# The Chaotic State of UK Drone Regulation

Scott McLachlan[1,2,4,6], Kudakwashe Dube[3,4], Burkhard Schafer[2], Anthony Gillespie[5], and Norman Fenton[1,7]

[1] Risk and Information Management, Queen Mary University of London, London, UK
[2] Edinburgh Law School, University of Edinburgh, Edinburgh, UK
[3] School of Mathematical and Computational Sciences, Massey University, Palmerston North, NZ
[4] Health informatics and Knowledge Engineering Research (HiKER) Group
[5] Electronic and Electrical Engineering, University College London, London, UK
[6] McLachlan Digital Ltd, London, UK
[7] Agena Ltd, Cambridge, UK

In December 2020 the law for drone pilots and unmanned aerial vehicle (UAV) use went into a transition phase in preparation for new EU international UAV regulation. That EU regulation comes into full effect as the transition periods defined in the United Kingdom's Civil Aviation Authority Air Policy CAP722 expire during December 2022 (CAA, 2020). However, international homologation regulation will not address the patchwork of inconsistent drone use regulations that exist in the United Kingdom from the layering of local and subordinate authority byelaws over UK aviation law. We provide an extensive review of local authority regulation of drone use on public open and green spaces, finding that many local authorities are unaware of the issues being created through: (i) inappropriately couched or poorly framed byelaws; (ii) multiple byelaws covering the same area by virtue of overlapping jurisdictions; or (iii) the lack of readily identifiable policies for drone use on public land. Overregulation, inconsistent regulation and regulatory disharmony are causing confusion for recreational drone enthusiasts such that it is never clear which public- or crown-owned open and green spaces they are allowed to, or prohibited from, flying. While the government and local authorities might like them to, drones are not going away. Therefore, we conclude, the easiest way to ensure citizens stay within the bounds of drone law that is intended to ensure public safety, is to make that law *comprehensible*, *consistent* and *easy to comply with*.

## 1. Introduction

Since the Wright Brothers inaugural flight at Kitty Hawk in December 1903, airplanes have amazed and captivated an ever-growing population of aviation enthusiasts. While this fascination can be seen before Kitty Hawk and the invention of model aircraft in the hobby of pigeon and dove racing that dates back to the fifth Egyptian Dynasty in 3000 BC (Allen, 2009), arguably the largest growth period for aviation enthusiasts has been during the jet age and certainly since the development of cheap and easy-to-use quadcopter drones. However, recreational use of drones has not always been without its issues.

Dramatic headlines promulgate fear for regulators and in the general public on two fronts. First, headlines that speak of the military's use of *killer drones* that enable a pilot seated safely at a computer in one part of the world to eliminate targets located in another are commonplace: with many inevitably spiralling into comparisons between autonomous drones, Skynet, and Schwarzenegger's *Terminator* robot (DailyMail, 2011; Hastings, 2012; McKnight, 2013; Sabbage, 2019; Wadhwa & Salkever, 2021). Second, while they seem far more plausible and may be grounded on actual incidents, there are headlines that exaggerate the potential of civilian drone use to intrude on our personal privacy and security: using claims that uncited cases are *happening everywhere* and creepy mock demonstrations to imply the privacy of the reader is already at risk (Rossin & Bomnin, 2018; Zhang, 2018). While there have been *peeping tom* drone allegations and a small number of prosecutions (Gogarty, 2017; Nauman, 2017), these have occurred far less frequently than the headlines would have you believe. As a result of hyperbolic headlines and Terminator comparisons, the distinction between large fixed-wing weaponised military unmanned aircraft and small consumer drones being used in the local green space or park by either a recreational aviation enthusiast or tech-loving teenager has been demonstrably lost. Small consumer drones are now also seen as weapons - assailing commercial airliners or encroaching on the sanctity of our private property. Alarming headlines, public money and considerable effort are being expended to either limit or completely shutter their general use ostensibly on the pretence of public protection. However, it is not certain whether public safety or privacy were ever truly at issue.

During the last several years many countries have enacted new civil aviation regulations targeted at unmanned aerial systems (UAS)[1], unmanned aerial vehicles (UAV)[2] and remotely piloted aircraft (RPA)[3]. However, these are fairly broad and technical terms that are intended to encompass the entire range of military, commercial and civilian unmanned aircraft that the general public and hyperbolic newspaper headlines usually refer to as *model aircraft* and *drones*. While the UK's Civil Aviation Authority (CAA) website seems to imply model aircraft and drones are considered synonymous[4], we believe the distinction between them is important: especially when it comes to developing and enforcing rules to regulate their use, and processes for assessing the attendant risks of using either type.

Several reviews have already aggregated and described drone regulation at the nation or state level, but these reviews could lead the reader to assume this top-level legislation is the only regulation that applies to day-to-day drone use. However, the majority of recreational drone use occurs on public property like open and green spaces, community parks, and public beaches whose use is regulated by local authorities. In England, local authorities include Borough, District and County Councils and Unitary Authorities. One previous work in 2019 identified that state law and local law within the UK may not be acting in harmony with respect to recreational drone use, and the author concluded that ambiguity and confusion result from this disharmony[5]. Our work investigates the current regulatory situation for drone use, and especially, recreational drone use on ostensibly public or crown land in England in order to ascertain whether recent regulation made in the intervening three years has resolved the disharmony and reduced ambiguity and hence, confusion. We do this through a review of national and local government regulations, byelaws and policies. No prior work was identified that reported comprehensively on these local authorities and their bylaws and, as such, this may be the first to review the patchwork mosaic of regulation that exists within a single country.

---

[1] UK CAA CAP722D defines UAS as: An unmanned aircraft and the equipment to control it remotely.

[2] UK CAA define UAV as: The unmanned aircraft.

[3] UK CAA CAP722D defines RPA as: An unmanned aircraft which is piloted from a remote pilot station.

[4] https://www.caa.co.uk/consumers/remotely-piloted-aircraft/our-role/an-introduction-to-remotely-piloted-aircraft-systems/

[5] https://www.juriosity.com/knowledge/article/fc542880-28db-43b8-90d9-9310f49f4778

Our process for identifying and reviewing local authority byelaws and policies for drone use consisted of:

1. Identifying a dataset of existing local authorities in England[6];
2. Identifying, reviewing and recording responses to individual January 2021 *Freedom of Information* (FOI) requests made to each English council seeking:
   a. Whether the council had any byelaws relating to the use of *unmanned aerial vehicles* (UAVs), or *drones*, from the lands under their control;
   b. If not, whether the council had any policies relating to the use of UAVs from their lands; and,
   c. Whether the byelaws or policies had been reviewed in relation to CAP722C as published by the Civil Aviation Authority in December 2020.
3. Manually reviewing any byelaws or policies provided with or referenced by councils in their FOI response; and,
4. Manually searching each council's web presence for:
   a. [(drone) or (UAV) or (model aircraft)] and [(byelaw) or (bylaw)]
   b. [(drone) or (UAV) or (model aircraft)] and [(policy) or (restriction)]
5. Where no byelaw or policy was identified at step 4, a search was conducted of each council's web presence for:
   a. [(drone) or (UAV) or (model aircraft)]
6. Where no byelaw or policy was located for the council, a final search was performed using Google[TM] for:
   a. ["council name"] and [(drone) or (UAV) or (model aircraft)]

Online web presence searches were used as they are analogous to the approach a recreational drone pilot might employ to identify whether flying is permitted on the open and green spaces within an authority's jurisdiction.

In addition to data collection and analysis of local authority byelaws, policies, and review of meeting minutes and other council-related artefacts that mention *drones*, the lead author also conducted unstructured discussions with drone instructors and drone pilots at several drone pilot training schools and flying clubs regarding their understanding and impression of previous, current (transitional) and upcoming drone regulation at both a national and local authority level.

The remainder of this work proceeds as follows. After a brief introduction to the history of model aircraft and drone technology it explores the current approach to risk assessment of airspace incidents involving drones. Then, after exploring drone regulation at the national level in the United Kingdom, it presents our approach to, and findings from, analysis of the policies and byelaws for drone use on public land within local authorities in England. It concludes with a discussion of the future for drones and drone use in England.

---

[6] This was provided from the publishing service of the UK government: https://assets.publishing.service.gov.uk/government/uploads/system/uploads/attachment_data/file/1026384/List_of_councils_in_England_2021.pdf and verified using the Local Government Association website: https://www.local.gov.uk

## 2. Drone Technology

Commercially available UAV primarily fall into two types: *fixed-wing* and *multi-rotor* (Boon et al, 2017). Each presents with different capabilities and limitations. Fixed-wing UAV have traditionally been known as *model aircraft* due to their similarity in both appearance and operation to real aircraft; in some cases even presenting as scale models of known aircraft types. With few exceptions and for at least five decades beginning in the 1930's, model aircraft have generally been powered by very small but noisy gasoline engines (Gudaitis, 1994). While mass-produced electric model aircraft were becoming available from the mid-1980's, it was not until the early 2000's that electric model aircraft would become inexpensive and commonplace, and through their much lower buy-in cost changed the hobby of remote control airplane flying by making it affordable to a much larger audience (Carpenter, undated).

The first UAV were undoubtedly military, and may have been explosive-laden hot air balloons raised by Austrian forces over Vienna in 1849 (Erceg, Erceg & Vasilj, 2017; Vyas, 2020). Control of *lighter -than -air* UAV was difficult and divested at the whim of the wind. It is said that while 200 of the Austrian balloons were launched, only 1 managed to find its target (Vyas, 2020). However, it wasn't until almost 50 years later when Nikola Tesla demonstrated his radio-controlled boat in 1898 that potential for a viable solution to the control challenge of UAV would be realised (Sellon, 1997).

The UK armed forces experimented with remotely controlled aircraft during the early 1900s (Prisacariu, 2017), but the 1939 American OQ-2 Radioplane was the first mass-produced drone (Custers, 2016). However, in both cases these UAV were designed and primarily used as target practice for anti-aircraft gunners (Custers, 2016; MilitaryFactory, 2017). The Japanese were also seen to use uncontrolled bomb carrier balloons during WW2 in much the same way as the Austrians described in the preceding paragraph almost 100 years earlier[7]. It wasn't until the Vietnam war that drones would fly combat surveillance and reconnaissance missions at scale and thus begin to mitigate risk to human pilots (Shaw, 2016).

While the UAV discussed in the preceding paragraph were fixed-wing unmanned aeroplanes, it was not until the late 1990's in Japan that the first quadcopter multirotor *drone* kits were developed (Darak, 2017). The first one to be produced in any real commercial quantities was the Canadian-manufactured Draganflyer 1 that became popular after it was used in the filming of Disney's 1999 movie *Inspector Gadget* starring Mathew Broderick (Darak, 2017; Draganfly, 2016). However, it was Frank Wang, CEO of well-known consumer drone manufacturer DJI, that revolutionised UAV technology by producing intelligent and easy-to-use *vertical take-off and landing* (VTOL) multi-rotor drones[8] and later, the stabilised accelerometer-driven camera gimbal. In 2012, DJI would produce the first robust, commercially available pre-assembled drone for enthusiasts and prosumers alike (Mac, 2015). Only two years after releasing that first Phantom drone, DJI were selling 400,000 units annually (Mac, 2015). Autel, Parrot and a range of smaller manufacturers have all since entered a retail market that was valued at over US$22.5Billion in 2020 (Wood, 2020).

---

[7] These were known as the Fu-Go Balloon Bomb. Around 9000 were said to have been released and they were identified flying or having exploded over Canada, America and Mexico. See: Mikesh, Robert C (2010). *Japan's World War II balloon bomb attacks on North America*. Smithsonian Institution Press. ISBN 978-0-87474-911-3. OCLC 745489144.

[8] Also known as *quadcopters*.

The distinction between *model aircraft* and *drones* is one of both temporal and functional context. Model aircraft have traditionally been the fixed wing scale simulacrum of aeroplanes and prior to the 1980's were almost exclusively powered by the combustion of gasoline or other inflammable substances. This can be seen in the regulation of the time, with authorities also regularly describing model aircraft as *powered by petrochemical vapours*[9]. The military have generally used UAV or *remotely piloted vehicles* (RPV)[10] to describe larger and significantly more sophisticated model aircraft they use in the theatre of war. *Drone* has been more commonly applied by the general public to describe the modern post-2000's multi-rotor UAV. Recent attempts in European and National regulation to merge the two terms ignores the vast differences between the two types of UAV and hence, their diverse prospective risk profiles. Examples include: (1) model aircraft, like their larger counterparts, require runway space for take-off and landing while multi-rotor drones can take-off and land from a stationary position (VTOL); (2) Model aircraft are generally considered less manoeuvrable than a multi-rotor drone in that model aircraft, again like their larger counterparts, must have airflow across the wing surface, and hence forward momentum, to maintain lift else they risk stalling and dropping from the sky. Conversely, the drone is capable of stopping, hovering, turning on the spot and instantly departing a location on any compass bearing.

## 3. Drone risk and safety assessment

There is no denying that a small number of potentially alarming incidents involving drones may have occurred[11]. We say *may have* because for most reported drone incidents by commercial pilots, a single pilot in a multi-crewed cockpit is the only witness. And even when a drone collision is claimed to have occurred, in many cases no damage has been found on the airliner[12] (Crumley, 2021). In the seemingly rare situations where a live strike collision has occurred and damage has resulted, it was later found to be professionally trained but less careful police drone pilots and not recreational drone enthusiasts who had caused the incident[13]. There are videos on the internet of drones being deliberately crashed at very high speed into decommissioned airplane wings[14], other videos purporting to be drone strikes on commercial airplanes that are actually Hollywood-style special effects[15], and articles with poor quality graphics suggesting drones could become ingested into jet engines with catastrophic effects[16]. However, these all seem geared more towards promoting fear of what could happen, and less about describing anything that actually is happening. The debate around whether one, more than one, or no drones at all were seen near Gatwick Airport in 2018 continues[17] (List, 2021; Shackle, 2020). However, the hyperbolic headlines proclaiming the dangers posed by the as-yet unverified one- to- two kilogram drone to aircraft weighing one-hundred tonnes or more[18] and airline passengers, whipped politicians and activists into a frenzy. As a result, some councils made parks and public open spaces no-fly zones on sometimes flimsy pretences[19] or simply as a result of negative discourse surrounding the Gatwick incident in the media.

---

[9] Seen later in the examples shown in Figures 6 and 7.

[10] With a preference for RPV in order to emphasise the fact that there was always a human pilot.

[11] These include when a drone at speed hit what is probably the smallest and least defensive helicopter available, the Robinson R-22, and the instructor and student pilot landed the craft safely and without injury or further incident - https://petapixel.com/2018/02/16/drone-causes-aircraft-crash-first-time-us-report/

[12] When an Envoy Air passenger jet's pilot claimed his plane had struck a drone 4 miles after take-off from Chicago O'Hare Airport, aside from the pilot's claim that it *might* have been a drone there was no corroborating evidence either of a drone strike on the plane itself, save the suggestion of a possible drone over a pond made by the pilot who took off immediately before the Envoy Jet. This created a possible suggestive condition, further fuelled by the fact that the aviation industry was on high alert following the Buttonville Ontario incident in footnote 13 several days before - https://dronedj.com/2021/08/24/envoy-air-passenger-plane-hits-drone-after-chicago-takeoff/

[13] Three days before the Envoy jet incident in footnote 12, a small Cessna 172 was struck by a York Regional Police drone at Buttonville Airport in August 2021. While dented and needing a replacement propeller, the Cessna landed safely. However, police only came forward to acknowledge the drone was theirs and admit fault after  local news media became involved. https://toronto.ctvnews.ca/plane-damaged-after-being-hit-by-york-police-drone-at-buttonville-airport-1.5554617

[14] https://www.youtube.com/watch?v=jEbRVNxL44c

[15] https://www.snopes.com/fact-check/drone-hit-plane-takeoff/

Risk assessments are regularly performed by safety regulators with a view to ensuring that products or systems are sufficiently safe for use. There are many methods and tools available for safety and risk assessment - and in the case of air proximity incidents tools such as Integrated Safety Assessment Model (ISAM), Fault Tree Analysis (FTA) and Event Sequence Diagrams (ESD) are commonly employed. In the United Kingdom (UK) the International Civil Aviation Organisation (ICAO) Safety Management System (SMS)[20] was previously used[21] to analyse aircraft proximity risk. The SMS risk assessment model describes a process wherein hazard identification, safety risk probability[22] and safety risk severity[23] are all used to identify risk on a two-dimensional scale ranked from A to E for severity, and 1 to 5 for probability[24]. Since at least 2019 the UK Airprox Board commenced using their own single-dimension risk ratings described in Table 1.

*Table 1: UK Airprox Board Risk Rating categories*

| A | Risk of collision: Aircraft proximity in which serious risk of collision has existed |
|---|---|
| B | Safety not assured: Aircraft proximity in which the safety of the aircraft may have been compromised |
| C | No risk of collision: Aircraft proximity in which no risk of collision has existed or risk was averted |

---

[16] https://www.911security.com/news/this-is-what-happens-when-a-drone-gets-sucked-into-a-jet-engine-video

[17] In fact, after expenditure of over £1million and claims of one witness seeing a drone, then another witness seeing two drones, then admission by witnesses that they may not have seen any drones at all, no evidence that a drone was ever near Gatwick airport that Christmas has ever been found. Police eventually admitted that there may not have ever been a drone, but that has not stopped the media hype, and politicians setting the Home Office (counter drone unit), NPCC (counter drone unit) and CAA on the course that has resulted in the Hostile Environment discussed in this paper.

[18] While the Airbus A320 mean take-off weight (MTOW) is around 77 tonnes, an Airbus A380 can run to 550 tonnes or more.

[19] For example: Within Ashford Borough Council the Addington and Bonnington Parish Council record in meeting minutes from the 14th October 2019 that they were looking to ban drones based on the fact that at their September meeting *a resident made reference to the flying of a drone on the playing field in an unsatisfactory manner*. Also; Within West Sussex, the Middleton-on-Sea Parish council who in meeting minutes on the 19th July, 2017 sought to put up signs banning drones on their large open space solely because *the neighbouring council had banned drones on their nearby open space and Middleton-On-Sea were worried drone flyers would move onto their open spaces.*

[20] The current version is: ICAO Safety Management Manual, Fourth Edition - 2018 (Doc 9859-AN/474). Found here: https://skybrary.aero/bookshelf/books/5863.pdf

[21] Many pre-2019 UK Airprox Report tables included a column called *ICAO Risk Rating.*

[22] On a scale from Extremely Improbable to Frequent.

[23] On a scale from Negligible to Catastrophic.

[24] As such a risk score is presented as a two-character value from A1 (Extremely Improbable and Catastrophic) to 5E (Frequent but Negligible).

| D | Risk not determined: Aircraft proximity in which insufficient information was available to determine the risk involved, or inconclusive or conflicting evidence precluded such determination. |
|---|---|
| E | Met the criteria for reporting but, by analysis, it was determined that normal procedures, safety standards and parameters pertained. |

# 4. Background to the regulation of drone technology

Australia claims to have been one of the first countries to regulate drones in their 2002 *Civil Aviation Safety Regulation* (Buchanan, 2016). By 2011, the Canadian-based *International Civil Aviation Organisation* (ICAO) were already considering the potential issues of, and proposing a regulatory framework for, drones[25]. The majority of countries began amending civil aviation rules to differentiate and distinguish recreational and commercial use of drones, and for many, to require registration of drones, licensing of some drone pilots, and to regulate allowed flight parameters like ceiling altitude and proximity to property and persons. Table 1 lists the relevant primary legislation and the year in which each was enacted or amended to regulate civilian and commercial drones for a select list of countries.

Drone legislation at the nation or state level tends to share commonality, in that most countries regulate metrics about the drone (weight), pilot (age), the approved flight ceiling (maximum altitude), minimum separation from property and people (minimum safe distance), minimum boundary around key facilities (no fly zones and controlled airspaces) and additional training, registration and licensing requirements for those who fly drones for commercial purposes. It is also common for these regulations to require drone pilots to secure permission from the landholders[26] of properties where they will take off, land, or overfly. For use on or over privately held property, this requirement had seemed to landowners and regulators to be entirely reasonable - it ensured the landowner had the opportunity to advise the drone pilot of potential issues or hazards[27] and overall was aware that the drone would be present, and reduced the likelihood of the drone pilot being summarily trespassed. However, it afforded private landowners an ability to restrict drone pilot access and drone use even over large *open access* land areas and potentially in conflict with the *right to roam*[28], only by virtue of the intention to use a drone. When applied to public or crown land (or common land[29]) the requirement to seek landowner permission has seen each public authority create their own approach for regulating drone use. This has resulted in a complex patchwork of confusing and sometimes contradictory permissions and hence, a *hostile environment* for drone pilots.

---

[25] Unmanned Aircraft Systems (UAS) (CIR328 AN/190)

[26] The CAA have removed this permission stipulation from the most recent version of the regulation; however it is still advisable to ensure you do not end up committing a trespass or cause alarm/disturbance for the landowner or occupants.

[27] Such as livestock, protected species of native fauna, or power lines in not easily seen places such as over the crest of hills or behind farm buildings.

[28] By virtue of the *Countryside and Rights of Way Act* 2000 any land shown on a map as open country, registered as common or dedicated open access land, or more than 600m above AMSL may be accessible under the right to roam. While the farmer is able to restrict ramblers accessing *open access* land for up to 28 days during the year, predominantly for lambing/calving season, or restrict the access of dog owners who let their dog run free around livestock or over moorland where game birds are bred and shot, they are not able to put up No Trespassing or other signs or barbed wire fences that would affect safety or deter the public from walking through access land.

[29] Common land may be owned by a local authority, privately, or by the National Trust. The *Countryside and Rights of Way Act* 2000 provides that there is usually a *right to roam* on property designated as common land. Conversely, the *right to roam* does not apply to village greens even though the local authority may allow grazing livestock, walking your dog or other forms of sport and recreation on them.

*Table 2: Introduction of primary Remote Piloted Aircraft (RPA) / UAV / Drone Legislation*

| Country | First UAV specific regulation | |
|---|---|---|
| **Australia** | 2002 | Civil Aviation Safety Regulation (CASR) |
| **Germany** | 2007 | Luftverkehrsgesetz [LuftVG] [Air Traffic Act] 2007 |
| | | Luftverkehrs-Ordnung [LuftVO] [Air Traffic Regulation] 2015 |
| **United Kingdom** | 2009 | Civil Aviation Act 1982, c. 16 |
| | | Air Navigation Order 2009 art. 138. |
| **China** | 2015 | Interim Provisions on Light and Small Unmanned Aircraft Operations (UAS Operation Provisions) |
| **Canada** | 2015 | Aeronautics Act, R.S.C.1985 |
| | | Canadian Aviation Regulations (CARs) |
| **New Zealand** | 2015 | Civil Aviation Act 1990 (as amended by CAA Docket 15/CAR/1) |
| **France** | 2016 | Order of December 17, 2015, Regarding the Use of Airspace by Unmanned Aircraft) (Airspace Order) |
| | | Order of December 17, 2015, Regarding the Creation of Unmanned Civil Aircraft, the Conditions of Their Use, and the Required Aptitudes of the Persons That Use Them) (Creation and Use Order) |
| **Israel** | 2016 | Aviation Law, 5771-2011 SEFER HAHUKIM [SH] [BOOK OF LAWS] (official gazette), 5771 No. 2296 p. 830 (as amended) |
| **Japan** | 2016 | Aviation Act, Act No. 231 of 1952 (amended by Bill No. 24) |
| | | House of Representatives (HR) Bill No. 24 of 189th Diet Session |

# 5. Drone regulation in the United Kingdom

In England, the Civil Aviation Authority (CAA) already administers drone use under provisions of the *Civil Aviation Act* 1982 and its subordinate regulation, the *Air Navigation Order* 2009, article 138. The key requirements for most drone pilots in the UK include:

- A Flyer ID to identify the pilot and demonstrate they have completed the basic 40 question online flying test.
- An Operator ID which must be affixed to each drone or model aircraft to identify the operator (be it company or individual) responsible for that aircraft.
- Rules that prescribe the category of drone and model aircraft operations, that are used both to identify the level of risk involved and class of drone that is permitted for use within that operational category (for example: A1, A2, A3, Specific, Certified).
- Rules that prescribe the class of drone pilots are allowed to use within each operational category (for example: Classes C0 - C4).
- The requirement for *visual line of sight* (VLOS) operation at all times for pilot-alone operations.
- The requirement for an observer who maintains VLOS when the pilot is flying using *first person view* (FPV).
- Maximum flight altitude.
- Minimum safe distance from people.
- Minimum safe distance from property.
- Requirement to check for Flying Restrictions and Hazards, including:
  - *Flight Restriction Zones* (FRZ) that include airports, spaceports, and other aircraft.

- *Restricted airspace* that include prisons, military installations, royal palaces and government buildings.
  - Events, emergency incidents, Sites of Special Scientific Interest (SSSI), tall structures, and other aircraft.
  - Notice to Airmen (NOTAMs), signs, apps and other resources that may have details of flight restrictions.
  - To be aware of local byelaws (as these may not be shown in drone pilot apps).

## 5.1 Drone regulation by local authorities

Delegation of State Powers to local self-government has existed since Roman times, and is common practice in many countries of the world (Fouracre et al, 1995; Serohina et al, 2019). Local governance by organisations such as territorial authorities, town trusts[30], corporations[31] and councils is believed to improve social cohesion by bringing the seat of government closer to the communities being governed and allowing citizens to be consulted on and involved in decisions on matters that directly affect them (Aulich, 2009; Khaile et al, 2021). The concept of local secular government in England began when the country was divided into *shires* and dates back at least to the 7th century Anglo-Saxons (Story, 2017; Williams, 1999).

Shown in Figure 1(a) and reminiscent of the borders of those Anglo-Saxon shires, are the largest local government bodies in England: the Counties. In Figure 1(b) we see that Counties are subdivided into District, Borough, Metropolitan and City councils that are generally responsible for provision of broader local services including: education, transport, household rubbish collection, recycling and housing[32]. It is common for County and District councils to receive a high degree of autonomy in administration of local affairs, with little interference from the State (Serohina et al, 2019). Figure 1(c) shows that within many District, Borough and Metropolitan councils are smaller parish, community and town councils that are typically involved in overseeing allotments, public clocks, bus shelters, community centres, parks and playgrounds[33].

---

[30] The first official local government in Australia was the Perth Town Trust, established in 1838.

[31] It was common in Australia for towns and cities to form a corporation. The Adelaide Corporation, precursor to Adelaide City Council, was created by the province of South Australia in 1840 - and was followed by the City of Melbourne and Sydney Corporations in 1842. In America, local governance is often performed under charter by Municipal Corporations. America also has a history of company towns - where one company owns many of the facilities and commercial establishments are owned by the same company who is often also the main employer.

[32] https://www.gov.uk/understand-how-your-council-works

[33] *Ibid.*

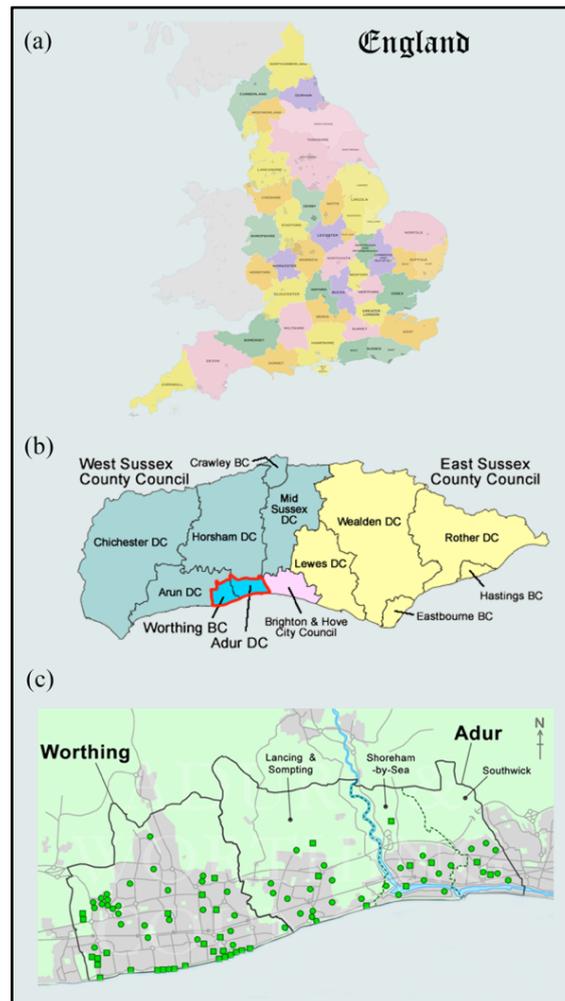

*Figure 1: Map showing (a) County Authorities of England; (b) District Authorities of West and East Sussex County Councils; and, (c) Parish Councils of Worthing Borough Council and Adur District Council showing parks with playgrounds (green circle) and open or green spaces and recreational reserves (green square)*

While the concept of *overlapping jurisdictions* is not new, more often they are discussed in the context of international trade and taxation, competition, contract, and intellectual property rights law (Mann, 2002; Stringham, 2006). Figure 2 portrays the three distinct hierarchies of regulatory bodies that exist over different open and green spaces in England, and demonstrates the various layers of public authority between the English Crown and Parliament and public open and green spaces. The aspect to note is that as you move closer to the apex of each triangle you move closer to the people, and the more likely it is that untrained citizens, or lay people, are involved in conceiving, formulating and developing the rules and regulations imposed upon the local community[34]. However, these self-government systems can be a double-edged sword. Many consider it resolved that advantages of local self-governance arise because the rules and regulations applying in a community have context, meaning and attachment to matters directly affecting the people who live within that community (Bailey & Elliot, 2009). However, this frequently ignores the potentially inevitable feature of municipal law making to deliver poor administration through misuse of local governance positions and delegated legislatory powers to empower personal agendas and greed, and the failure to maintain law and order in anything alike a fair and even-handed manner; which are both a matter of record (Cason, 2012; Edgar, 2016; Pasquini et al, 2013; Vibert, 2007). It also disregards the potential for poorly motivated, flawed or hurriedly developed local byelaws, allowed as delegated regulation by an Act of Parliament, to contradict or overregulate some common activity[35]. As the discussion above showed, local legislators face a formidable challenge. They need technical expertise regarding drones, drone safety and pilot training and certification, and a sound understanding of the subject of regulation; expertise they will often lack. But similarly, they also need an appreciation of how to draft rules in an environment where (a) drones can fly across regulatory boundaries and (b) knowledge of how their local laws interact horizontally and vertically with other rules, including national law. We noted above that the issue of overlapping jurisdictions is normally encountered in 'big ticket' legislation like international copyright law – where despite the availability of expertise, professional civil servants and the ability to draw in external experts rules still often clash. However, for local authorities this may represent an almost insurmountable task, and we will see below that it does indeed frequently result in low quality regulation.

---

[34] This also has a long and interesting history. While in medieval times the townsfolk tended to be responsible not just for their own behaviour, but also the behaviour of their peers, the more powerful and better off in a community felt they should be seen as responsible for dealing with local crime and crime prevention - which led to formation of the Justices of the Peace. Thus, a system where everyone was involved and responsible in deciding what was a crime and who was responsible gave way to one where this was the job of certain, often prominent and well-off, people.

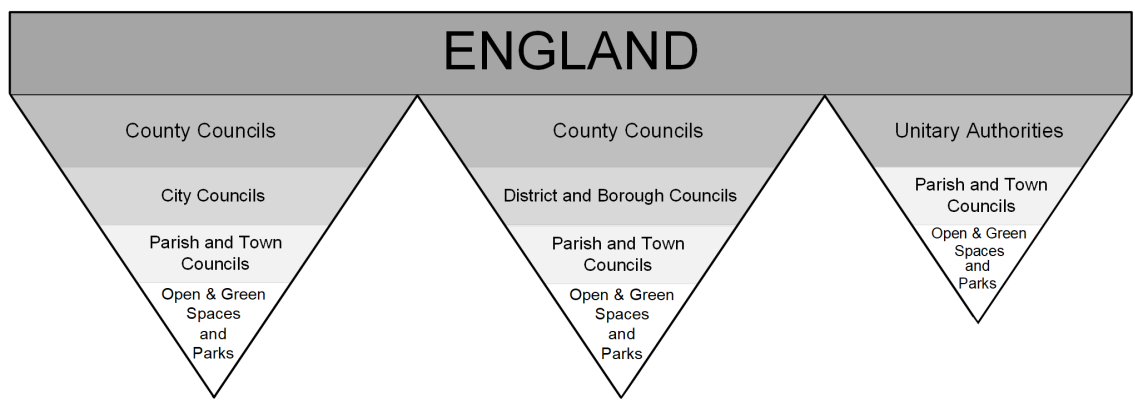

*Figure 2: Inverted Pyramids of Governance and Authority Ranking*

Council boards have tended to be populated in two ways. For some, councillors are voted into office by the wider community in local government elections; while for others it is a quorum of party or sitting members who elect new members (Pasquini et al, 2013). In the first case the potential new council member becomes known to the community while campaigning for votes. However, the process for identifying people who might be appointed in the second can be far less open and accountable. While examples of documents describing processes for currently sitting member-appointed councils describe identifying a *qualified person*, the qualifications may only be as strenuous as *familiarity with the processes of local government* or potentially *someone who will "go along" with the majority of the board*, or as flexible as someone with the right *political alignment* or even *who is known to be available on meeting nights* (Angerer, 2011; Pasquini et al, 2013).

The requirement to *be aware of local byelaws* is perhaps the most, if not the only real obstacle contained within existing UK CAA drone regulations. This is because it is through the use of local authority byelaws that the existing *hostile environment* for drone enthusiasts arises. As shown in Table 3, 219 local authorities (71% of all 310 councils reviewed in this work) either do not have a byelaw or policy for drone use from their open or green spaces, or if they do, these byelaws and policies were not readily identifiable or accessible from their online public presence. Figure 3 uses a subset of 100 councils drawn from the complete dataset of 310 described in Table 2. This subset consists of all county councils with the addition of the largest of the metropolitan and unitary authorities. In that group only 26 have easily identifiable drone policies: demonstrating that drone policies for the larger councils are the exception rather than the rule.

---

[35] For example, in *Strictland v Hayes Borough Council* (1986) where a bylaw prohibiting singing or reciting of any obscene language generally was held to be unreasonable and as a result, the passing of this delegated legislation was found to be *ultra vires* and rejected.

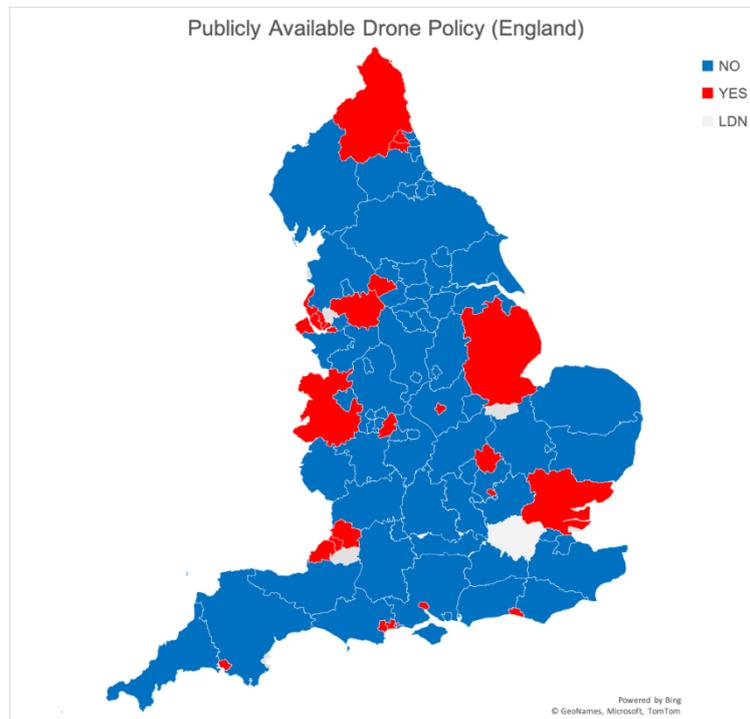

*Figure 3: Publicly available drone policy - subset of 100 councils*

While most councils have no publicly identifiable drone policy, within their borders many have *open or green spaces* operated by the council, public committees or parish councils that do. Figure 4 uses the same subset of councils as Figure 3 to show the disposition of the accumulation of council, parish council, and *open and green space* policies on drone use. When we consider that Figure 4 only represents a little less than one third of all councils reviewed in this work and consider that there are almost 200 smaller district councils not included in this visualisation, we start to understand the confusing patchwork of drone regulation that functions as the most local layer of the *hostile environment* in England.

Table 3: Summary statistics: Local Authority drone byelaws and policies - England

| | Number | Publicly available drone policy | REC - Blanket Ban (no exceptions) | REC - Blanket Ban (site specific allowances) | REC - Restricted (must apply for permit) | REC - Permitted (reference to CAA rules) | REC - Permission (site specific prohibitions) | REC - No blanket policy (site specific prohibitions) | COM - Blanket Ban (no exceptions) | COM - Blanket Ban (site specific allowances) | COM - Restricted (must apply for permit) | COM - Permitted (reference to CAA rules) | COM - Permission (site specific prohibitions) | COM - No blanket policy (site specific prohibitions) | Additional permit for filming with drones | Permit application fees apply | No identifiable drone policy |
|---|---|---|---|---|---|---|---|---|---|---|---|---|---|---|---|---|---|
| **District Councils** | **193** | 57 | 36 | 9 | 21 | 21 | 12 | 6 | 24 | 7 | 41 | 17 | 12 | 4 | 16 | 2 | 136 |
| **County Councils** | **26** | 2 | 7 | 0 | 2 | 4 | 1 | 3 | 2 | 0 | 7 | 4 | 1 | 3 | 3 | 1 | 24 |
| **Unitary Authorities** | **55** | 17 | 9 | 1 | 9 | 6 | 1 | 2 | 2 | 0 | 18 | 7 | 1 | 1 | 17 | 5 | 38 |
| **Metropolitan Councils** | **36** | 15 | 4 | 0 | 6 | 2 | 5 | 1 | 1 | 0 | 9 | 1 | 5 | 0 | 4 | 2 | 21 |
| **Totals** | **310** | **91** | **56** | **10** | **38** | **33** | **19** | **12** | **29** | **7** | **75** | **29** | **19** | **8** | **40** | **10** | **219** |

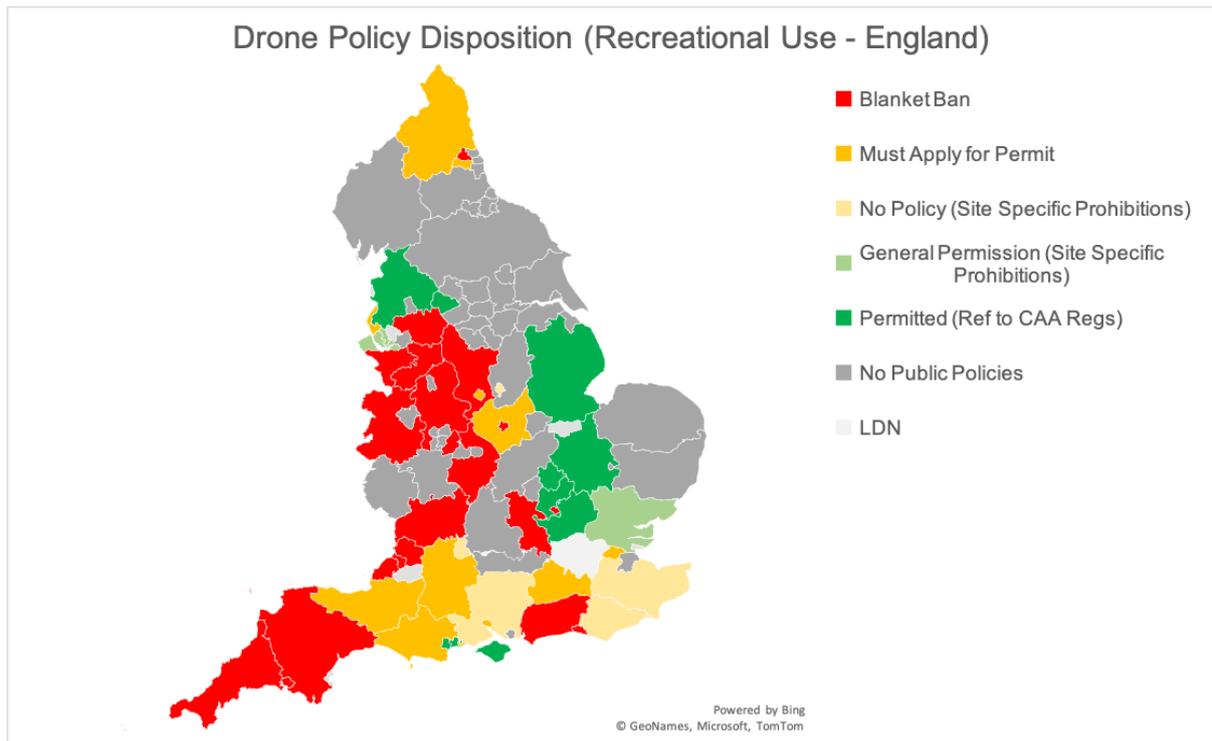

*Figure 4: Disposition of Drone Policies identified in Council Regions - Subset of 100 councils*

Of the 77 councils in our complete dataset of 310 that did have a publicly identifiable policy, 38 (49%) require even the recreational drone pilot to: (a) apply for a permit; and (b) have between £1million and £10million[36] in public liability drone insurance before flying[37]. We identified that 56 (73%) have a blanket ban on all recreational drone flying within their jurisdictions, and that 10 (13%) have a general ban with some limited and controlled sites where drone flying is permitted[38].

For the hobbyist or recreational drone enthusiast who has a small drone that he or she takes out with their child/ren, and that by virtue of weighing over 250grams[39] or having a small camera[40] falls within consideration of the overriding EU law[41], the CAA regulations and bylaws, this effectively means that: in almost half of England's local authority districts they remain unsure as to where and whether in fact they are allowed to use their drone; in 104 (33%) they are expressly (n=66) or effectively (n=38) banned; and while there is no prohibition at the council level for 31 (10%), there are site specific prohibitions on some or most of the open and green spaces. Currently, we could only confirm unrestricted permission to fly drones from the open or green spaces in just 33 (11%) of the council jurisdictions reviewed in this work.

---

[36] East Devon District Council were at the extreme end of this insurance requirement, with drone pilots required to demonstrate proof of a £10million indemnity policy for drone use.

[37] Many drone pilots the lead author spoke to when visiting drone clubs consider the requirement for an application for a permit each time the pilot wishes to fly to recreational drone to be nothing more than a *ban by paperwork* - that the council don't want to be cast in a negative light by saying "*you can't do this*", so they wrap a formal paperwork process around it knowing very few recreational pilots will apply and fewer still will receive permission so that on the surface it ostensibly looks like they are saying "*you might be able to do this.*" Completing application for permit forms and paying for a high-value indemnity insurance policy - especially one sufficient to allowing you to fly in any council jurisdiction - in order to then get to fly one or a few days seems overburdensome - especially when your ability to fly on that day, if such a permit is approved, remains at the whim *inter alia* of the weather.

[38] These sites tended to be places that had previously been annexed as model aircraft aerodromes or where a local flying club maintained a permit for use.

[39] While writing this work we spoke with drone enthusiasts at several flying clubs. They report that drones weighing under the 250gram minimum weight and falling into the 'Toy' class such as to be considered outside many of the requirements of existing legislation and most bylaws, were either rare or, as several people were happy to demonstrate, so light as to be unflyable in anything less than perfect and still outdoor settings - because even the slightest puff of wind would deviate the drone's course. Several people described an outdoors-flyable 'Toy' class drone under 250grams that also falls outside of these regulations as a *unicorn drone*.

[40] The vast majority of small drones have cameras, and the presence of a camera uplifts the drone out of consideration of the 'Toy' class. Even the 242gram DJI Mini2 that falls just under the 250gm weight class limit allowed for 'Toy' classification has a camera, meaning it does not fall within the 'Toy' class.

However, the overall state of confusion doesn't just exist for the drone pilot. Of the 193 district councils, 19 in their FOI responses spoke of absolute prohibitions to drone use, annexing outdated public spaces byelaws enacted between 1950 and 1990 like those shown in Figure 5 and Figure 6 that regulate the use of *model aircraft* driven by *combustible substances*[42]. Whether ageing provisions intended to cover noisy, smoke-belching model aircraft fuelled by inflammable liquids that in a crash may ignite fires, as the FOI respondents seem to contend, would also cover the use of modern battery-powered drones that are significantly quieter and more environmentally friendly, is debatable.

13. A person shall not fly any power-driven model aircraft, that is to say, any model aircraft driven by the combustion of petrol-vapour or other combustible substances, in the pleasure ground.

Figure 5: Section 13 from the 1963 Borough of Andover byelaws provided by Test Valley Borough Council[43]

"model aircraft" means an aircraft which either weighs not more than 5 kilograms without its fuel or is for the time being exempted (as a model aircraft) from the provisions of the Air Navigation Order;

"power-driven" means driven by the combustion of petrol vapour or other combustible vapour or other combustible substances.

**Model Aircraft**

2. No person on the Common shall release any power-driven model aircraft for flight or control the flight of such an aircraft.

3. No person shall cause any power-driven model aircraft to take off or land on the Common.

Figure 6: Sections 1-3 from the Reigate Heath Byelaws provided by Reigate and Banstead Borough Council[44]

As shown in Figure 7 other councils[45] have amended their definitions for model aircraft to incorporate those with *one or more electric motors*; which at least one council, Runnymede Borough Council, recognised had not been tested in a court of law and hence, *may not legally cover UAV's*[46].

---

[41] Recital 16 specifically excludes cameras on units classified as 'toys' due to the perceived risk to privacy and the need to protect personal data.

[42] For example: Adur and Worthing Councils, Medway Council, North East Lincolnshire Council, Reading Borough Council, Test Valley Borough Council, Reigate and Banstead Borough Council

[43] https://www.whatdotheyknow.com/request/byelaws_relating_to_uav_flights_678

[44] https://www.whatdotheyknow.com/request/byelaws_relating_to_uav_flights_639

[45] For example: Runnymede Borough Council, South Ribble Borough Council, North West Leicestershire District Council, Gedling Borough Council

[46] Quoted from FOI response letter of Runnymede Borough Council dated 3 March 2022. Sourced from: https://www.whatdotheyknow.com/request/byelaws_relating_to_uav_flights_646

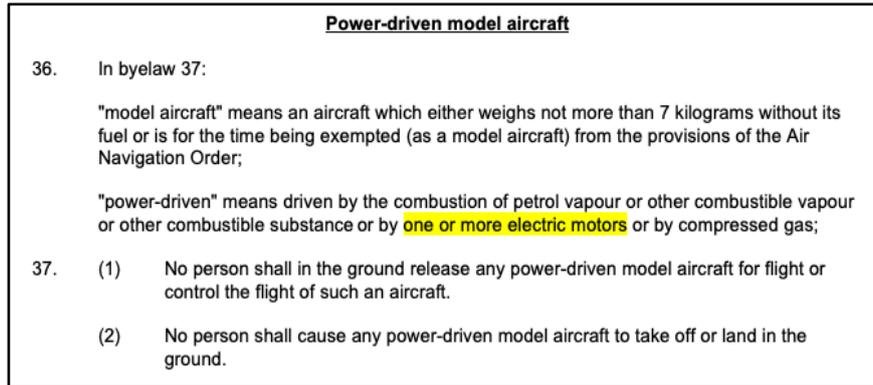

*Figure 7: Sections 36-37 of the Runnymede Borough Council's Pleasure Grounds and Open Spaces Byelaw[47]*

Some councils have inconsistent and poorly framed byelaws across different green or open spaces within their jurisdiction, and that their FOI responses suggested were prohibitions on drone use. For example, Guildford Borough Council in their FOI response maintained that council byelaws regulate model aircraft use[48]. However, on reviewing the byelaws on their website it was, as shown in Figure 8 we identified that whilst the word *unmanned* had been added to one byelaw that regulates three large common spaces, this byelaw would not actually cover most drones as they commonly weigh less than 5kg.

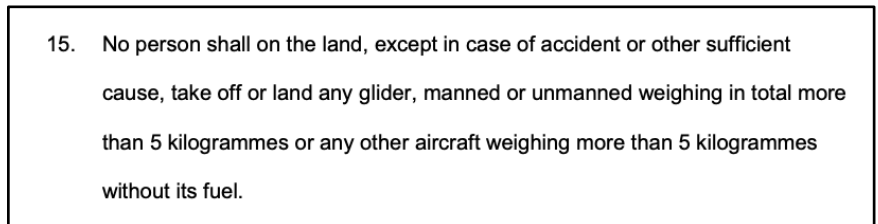

*Figure 8: Section 15 of the West Heath, Pirbright Common and Bullswater Common Byelaws provided by Guildford Borough Council[49]*

Further, the contextual framing of Guildford Borough Council's other open space byelaws, such as that shown in Figure 9 for Merrow Downs, prescribe a situation where the contravening person is *in* the aircraft that is landing or departing from the council-managed public site. Again, this would not cover drones as they are generally unmanned.

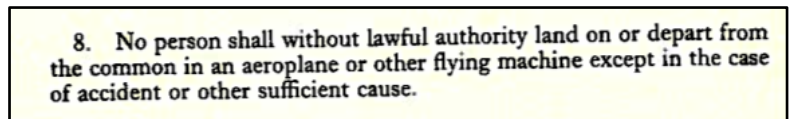

*Figure 9: Section 8 of the Merrow Downs Byelaw as provided by Guildford Borough Council[50]*

At least 36 councils also required an application for a filming permit prior to flying any drone with a camera. Some required this application be made to the filming office of the larger shire county district, for example - Ashford Borough Council in Kent required a filming permit from the Kent Film Office. It may be that other councils within the same district, for example - Canterbury City Council, also require an application to the Kent Film Office but because there

---

was no identifiable drone policy on their website advising of this requirement, the recreational drone pilot would be unaware.

There were also examples of what this work terms *disharmonious regulation overlap*. These were situations where a parent local authority had decided a policy either in favour of or against drone use[51], and a subordinate local authority within their area of control had rendered a policy to the opposite effect. Often, the subordinate authority was either a parish council or a committee overseeing large open spaces, moors, beaches or sporting fields[52].

It could be contended that in this instance the most appropriate solution would be regulation of drone use on public- or crown-owned open and green spaces at the UK level. This would serve multiple purposes, including: providing consistent country-wide drone use regulation and enforcement; reducing the requirement and costs for several hundred councils, and several thousand parish councils and other subordinate authorities, that must review, understand and then make decisions about drone use in their areas; reduce the current incidence of disharmonious regulation overlap; and finally, it would ensure recreational pilots can be more aware of where they are allowed to, or prohibited from, flying.

Thus far we have not considered whether byelaws that operate as a blanket ban of drones in all open and green spaces across an entire council jurisdiction are valid. Here we ask the question: *can a local authority blanket ban all use of drones from pleasure grounds, open spaces and public walkways within their borders?*

Councils are empowered to make byelaws[53] for the purposes of good rule and government, and for the suppression of nuisances, across the whole or any part of the area within their boundaries[54]. Aside from the *Local Government Act* 1972 (LGA), other legislation may also enable the making of byelaws to regulate pleasure grounds, open spaces and public walkways[55]. Certainly, the local authority's ability to make byelaws to regulate public spaces depends on two things: *firstly*, the local authority's interest in the land; and *secondly*, the local authority's involvement in the management of the land.

A local authority byelaw can be invalid where it: (i) regulates matters outside of the powers or authorising statute of the authority making the byelaw; (ii) lacks certainty and clarity as to what action is required or prohibited; (iii) is inconsistent with or prohibits something that is allowed by existing law[56] (also see Figure 10); or (iv) is manifestly unreasonable, not in good faith or was made on incorrect grounds[57].

---

[51] Or, in response to their FOI request, they replied that the council had no byelaws or policies prohibiting drone flying from the public lands they control and/or referencing to CAA regulations to say that drone use within those regulations was likely permitted.

[52] For example: Chorley Borough Council, East Hampshire District Council, Eastleigh Borough Council, Hastings Borough Council, Cambridgeshire County Council.

[53] *Local Government Act* 1972, Section 236.

[54] *Ibid*, Section 235.

[55] Including: *Public Health Act* 1875 and *Open Spaces Act* 1906. For a list of matters smaller local authorities like Parish Councils have the power to make byelaws for, see Appendix B.

[56] 47 at Section 235(3):
> Byelaws shall not be made under this section for any purpose as respects any area if provision for that purpose as respects that area is made by, or is or may be made under, any other enactment.

[57] The Wednesbury unreasonableness test.

Local authority byelaws are subject to the scrutiny of the courts in two ways: (i) by way of direct challenge to and judicial review of a decision made under that byelaw[58]; and (ii) a collateral challenge raised as a defence when being prosecuted *under* that byelaw, that challenges validity *of* the byelaw.

> Byelaws are considered measures of last resort after a local council has tried to address the local issue the byelaw applies to through other means. A byelaw cannot be made where alternative legislative measures already exist that could be used to address the problem. Byelaws should always be proportionate and reasonable. Where a byelaw is no longer necessary, it should be revoked.

*Figure 10: Extract from GOV.UK[59]*

Many of the local authorities reviewed in this work have policies which, as discussed earlier, would seem to require drone pilots to apply to the local authority itself, or another organisation, and pay a fee simply for *permission to film or photograph* within that council's boundaries. Examples include the Port of London Authority (PLA) who choose to justify the use of drones when a fee has been paid, and prohibit their use when it has not[60], and Oxfordshire Country Council, West Sussex County Council, Buckingham Shire County Council, East Devon County Council and others who similarly all require a fee for a permit to use a drone that has the capability to film with an onboard camera. As barrister Richard Ryan points out[61], by relying on erroneous byelaws that misrepresent the legal position of the CAA, local authorities create the impression that they are happy to grant TV companies like the BBC permission, possibly because TV is seen as free advertising, while unreasonably preventing legitimate drone users from filming - even where filming is only a by-product of using the drone that came equipped from manufacture with a camera. The CAA consulted with legal counsel and weighed in on this issue in October 2018[62], rendering a decision on the ability of an authority, in this case the PLA mentioned above, to demand that drone pilots apply for and pay fees to receive a permit to fly over or near the area they control. They found that where the flight is initiated from and culminates on land[63], an application to and permission from the PLA *is not required*. DroneSafe UK also note that there is a strong argument to be made that the requirement for PLA permission to film over the river was also unnecessary, given that during a safe flight that remains within the CAA regulations, the drone is required to remain at least 50 metres from vessels and people on the river[64]. DroneSafe UK ultimately acknowledge that there may be some resistance from the PLA regarding this decision given that the PLA *stand to gain significant revenue if they can force all pilots to apply and pay for flight permissions*. It should be noted that policies and byelaws requiring application (often with a fee) for a permit to film as a proxy for banning or limiting drone use are yet to be tested in a court of law, and for the same reasons it is possible they would be found unenforceable.

---

[58] By operation of Part 54 of the Civil Procedure Rules.

[59] https://www.gov.uk/guidance/local-government-legislation-byelaws

[60] http://www.pla.co.uk/Safety/Use-of-drones/unmanned-aerial-vehicles-UAVs

[61] https://www.juriosity.com/knowledge/article/fc542880-28db-43b8-90d9-9310f49f4778

[62] https://dronesaferegister.org.uk/blog/caa-finally-provide-clarification-on-uav-flights-over-the-river-thames

[63] Drone fights taking off from and landing on boats on the Thames would seem to be within the PLA's jurisdiction and have resulted in heavy fines as a result of regulations, for example, that restrict the release or disposal of objects from vessels on the river.

[64] Where the PLA would naturally be involved would be in situations where traffic along the river would be affected by the proposed drone flight - such as situations where the river would need to be closed or diverted around the area where the drone is working. An example of this might be if the drone was being flown for an extended period under the full length of London Bridge in order to develop complete photogrammetry images of the structure as part of planning structural restoration or rectification works.

Wednesbury unreasonableness is the standard used when assessing an application for judicial review of a public authority's decision. A reasoning or decision is Wednesbury unreasonable (or irrational[65]) if it is so unreasonable that no reasonable person acting reasonably could have made it[66]. The test has two limbs (Parsons, 2020). The *first limb* focuses on the decision-making process of the local authority and has an emphasis on whether that authority has taken into account the right issues when reaching their decision. The *second limb* focuses on the outcome and asks whether, *even if the right things have been taken into account*, the resulting decision is so outrageous that no reasonable decision-maker could have reached it. For those authorities who have enacted byelaws that create a blanket ban on recreational (and in some cases, commercial) drone use, a court would have regard to *whether it is reasonable to enact a blanket ban on an activity,* like drone use, *within an entire local authority area*. It is highly likely that banning drones in their entirety *would not* be considered reasonable. However, banning *certain* drones such as those over a *certain* weight in *certain* locations *might* be considered more reasonable because it narrows the focus and does not unjustly constitute a blanket ban. If byelaws establishing blanket bans were found to be unreasonable, the court would also find the byelaw to be invalid and therefore not enforceable.

In any event, where it concerns drones the Civil Aviation Authority (CAA) are the sole regulator for UK airspace. As regulator, the CAA have already acted to regulate the use of drones in the UK. To that end and in regards to some aspects of the matters discussed in this work, it may be possible that a court could find local authorities have acted in conflict with existing regulation where they have created byelaws to completely ban or heavily regulate a matter already addressed by existing CAA regulation.

## 5.2 New national drone regulation

In December 2022 the requirements of *Commission Implementing Regulation (EU) 2019/947 of 24 May 2019*[67] comes into full effect as retained EU law contained within flight regulation CAP722[68]. Much of the EU regulation is intended towards identification of potential safety issues, and classification and mitigation of risk. However, while it repeats the work *risk* more than 40 times within the document, it focuses more on drone and pilot registration, describing *classes* to classify drone use[69], and existing and new data that must be collected by pilots and regulators: *paperwork* in the form of registration data and reports. The one section with *safety* in the title (*Article 19: Safety Information*) consists of five points in less than half a page focusing on the requirement for regulators to have standardised processes for reporting and assessment of potential safety issues, and the section on risk assessment (*Article 11: Rules for conducting an operational risk assessment)* is prospective in context but written predominately in legalese and not generally approachable to the average recreational drone pilot. They appear intended for directing regulators in development of an approach for risk assessment in their description of prospective hazards that may impact the safety of a drone flight[70]. Yet, when one looks through the regulation for guidance on how these risks should be assessed and scored, it is silent. It is only when we turn to the related Opinion 05/2019 document (EASA, 2019) that we find some description of the process in the form of process steps with static two-dimensional tables for risk classification and assessment. However, several of the steps and the tables they contain use the drone's weight, width, or potential maximum kinetic energy it might impart in a collision in some cases as proxies for more causally important human and environmental factors[71].

---

[65] In *Council of Civil Service Unions v Minister for the Civil Service* [1985] AC 374, [1984] 3 All ER 935 Lord Diplock called this 'irrationality' and he went to say at 410: 'By "irrationality" I mean what can by now be succinctly referred to as "*Wednesbury* unreasonableness". … It applies to a decision which is so outrageous in its defiance of logic or of accepted moral standards that no sensible person who had applied his mind to the question to be decided could have arrived at it.'

[66] *Associated Provincial Picture Houses Ltd v Wednesbury Corporation* (1948) 1 KB 223

[67] https://www.easa.europa.eu/document-library/regulations/commission-implementing-regulation-eu-2019947

[68] https://publicapps.caa.co.uk/docs/33/CAP722%20Edition8(p).pdf

[69] Open, Specific, Certified

There remains a distinct need for regulatory harmonisation between UK Law and local authority regulations. However, adopting international homologation similar to how motor vehicles are certified for conformity has not delivered this. Adopting the EU drone regulation has vastly increased complexity while failing to resolve any of the UK's internal regulatory inconsistencies. Withdrawal from the EU law places the UK in a complicated and unenviable position. Much EU law is and remains absorbed within our domestic regulation, yet the continued presence of that EU regulation in our domestic law does not guarantee automatic cross-country compatibility or consistency. One of the primary aims of adopting the new EU drone laws in the UK was to harmonise drone operations across the entire EU airspace - something media articles lauded as *streamlining* and *beneficial for drone pilots*[72]. However, our withdrawal from the EU changed that landscape. Unlike the situation for most EU countries who are adopting these new regulations, qualifications and permissions obtained in the UK will not always be valid in EU member states[73]. This means that UK drone operators wishing to fly in most EU countries and EASA associate member states[74] will be required to register in each member state they wish to fly in and, depending on whether the UK qualifications[75] have become acceptable in that state, will almost certainly be required to complete the local equivalent to their UK qualification. The same is true for EU drone pilots coming to the UK. They will need to register with the UK CAA and complete UK versions of the drone pilot qualifications to fly within UK borders. Therefore, the efficiencies and benefits of this cross-border harmonised regulation are presently only realisable by drone pilots in EU member states who seek to fly in other EU member states - for example, a drone pilot qualified and certified in Germany would be able to use the same certification in France, Denmark or Italy without this additional retraining and expense. While the UK could request recognition of our CAA certifications in EU member states, the EASA report our government and aviation regulator have yet to initiate that process[76].

The European Union Aviation Safety Agency (EASA) admits, much of the data for ground based risks was based on anecdotal stories and hence is unsupported by evidence[77] (EASA, 2018). The same EASA document implies that reports of drone incidents by airline pilots have been considered resolutely factual[78]. However, the research conducted for this work found that data supporting this inference is lacking and airline pilot's accounts should also largely be characterised as anecdotal. For example: some reports describe the observations of only a single pilot in a multi-pilot cockpit[79] - the other pilot not having seen the *drone*. Others have resulted in headlines that speak of *serious near-misses* (Tonkin, 2017) while indicating that civilian drones were observed at altitudes[80] or in weather conditions[81] far exceeding the flight tolerances and abilities of even quite expensive multi-rotor consumer drones of the day. While the Airprox Board consistently accepted pilot's reports as fact and found a significant level of risk to safety existed due to the presence of an alleged and often unverified drone, there was almost never any evidence to substantiate these findings[82].

---

[70] Including personnel competencies and experience, airspace volume, the density of human population and other potential hazards in the area being overflown, proposed flight metrics including altitude and the airspace classification, and weather conditions.

[71] For example: Human Factors (training, experience, decision making ability etc.) and prevailing weather conditions at the time of the flight.

[72] https://www.bbc.co.uk/news/technology-55424729

[73] https://www.heliguy.com/blogs/posts/brexit-impacts-new-drone-laws

[74] Iceland, Liechtenstein, Switzerland and Norway.

[75] A2 CofC and GVC.

[76] https://www.easa.europa.eu/faq/123767

[77] Section 2.1 of Opinion No 01/2018 by the EASA regarding their proposed regulation of all civilian unmanned aerial systems (https://www.easa.europa.eu/document-library/opinions/opinion-012018)

Pilots and air traffic controllers are required to report[83]: (i) collisions to the Air Accidents Investigation Branch (AAIB); and (ii) aircraft proximity incidents that have potential to compromise safety to the UK Airprox Board (Airprox). In twelve of 2017's twenty-two reported incidents the Airprox Board speak of the airline pilot's *inability to avoid the [drone]*, yet it is notable that not a single collision occurred - meaning that not only *did* the pilot avoid the drone but incredibly and in almost every case, he or she did so without affecting the aircraft's heading or speed. This would support the assertion that commercial airliners in most Airprox reports were not at risk to the degree adjudicated in the Airprox Board's risk assessments[84]. While there are millions of consumer drones in use today, contrary to the claims of airspace regulators, some academics and the mainstream media - there have been no recorded incidents of a drone actually colliding with or causing serious damage to a large commercial airliner or its passengers[85]. This informs our prior belief that such serious and life-threatening incidents have a very low likelihood of occurring. When risk scoring an individual incident report we should update that prior belief using observations that directly inform two hypotheses: First, *that a drone was present*; and second, *that a collision between that drone and the reporting airliner was likely to occur*. In order to arrive at a decision about each hypothesis we should weigh the reported evidence, including: (a) the likelihood that the object observed was a drone based on the altitude and other environmental conditions at the time; (b) whether both or only one of the pilots in the cockpit of the aircraft observed and reported the object; (c) whether the observing pilot found it necessary to perform an emergency course adjustment to deviate away from the object; and (d) whether there is corroborating evidence either from airplane or ground systems or other independent sources[86].

---

[78] *Ibid*, under the heading **Safety Issues**.

[79] E.g.: Airprox Report 2017056 - A helicopter pilot reported a close fly-by of a drone that he claims he had to *swerve to avoid*, and which came *within 20-40ft of the front left side of his craft*. His co-pilot did not report seeing the drone. This fact scenario is typical of many Airprox reports reviewed during this work.

[80] E.g.: Airprox Report 2017035 - On 2nd February 2017 the captain of Airbus A319 flying from Leeds to Edinburgh claimed to have 'almost crashed into' a civilian quadrotor drone that he saw for only 2 seconds flying above him at an altitude of 18,200 feet. 18,000 feet converts to about 5,500 metres above the ground - beyond the reach of many civilian quadcopter drones (for reference the DJI Inspire maximum altitude is 13,100ft and the DJI Mavic Air maximum altitude is 16,404 feet).

[81] E.g.: Airprox Report 2016213 - On 1st October 2016 the captain and co-pilot of a Boeing 737 claimed to have come within 30 metres of a *50-100 centimetre wide* red and black drone flying off their left wingtip at an altitude of 6000ft (1800 metres). The Boeing was negotiating 'severe weather' of a type that most captains were diverting to avoid and that had 'flooded the runway', making it unlikely that a small drone would even be capable of anything approximating controlled flight at that altitude and speed. It is more probable that they saw their own cockpit lights or some other such phenomena reflected in the windows of the cockpit against the dark thunderclouds outside, or another airliner that was much farther away (and hence looked smaller due to perspective). Consumer quadcopter drones that do come in the *red and black* colours described include the DJI Mavic (Flame Red) and Parrot Bebop (Red), neither of which could fly in the weather conditions described in the airprox report.

[82] Of the 22 incidents listed in the Airprox 2017 consolidated drone report, only one was made by a drone pilot. 21 were made by military, airline and commercial helicopter pilots and of those, only two - 2017067 & 2017033 - were dismissed as providing insufficient information to make a determination; even though in the latter incident both pilots corroborated the sighting. In every other case (n=19), many providing less detail than the two that were dismissed, the Airprox Board found that *safety had not been assured* (n=8) or *safety was compromised* (n=6), or that *definite risk of a collision had existed* (n=5). The high risk ratings given by the Board belied the innocuous descriptions provided and the fact that for many only a single pilot in a multi-pilot cockpit observed the *alleged drone*. Many (n=12) also spoke of the pilots *inability to avoid the [drone]* even though there had been no single reported collision with any of the alleged drones.

[83] Most reporting is performed using the centralised EU reg 376/2014 compliant ECCAIRS 2 system at https://aviationreporting.eu

[84] An assertion shared by the authors of the Airprox Reality Check website (https://airproxrealitycheck.org) who, utilising their own Reality Check Process, re-evaluate Airprox reports and demonstrate that there are other potentially more credible explanations than drone incursion.

Let us consider a fact scenario where: (a) the airliner was at 18,000 feet and beyond the maximum altitude of many of even the most capable consumer drones; (b) only one of the two pilots in the cockpit observed the object claimed to be a drone and for a duration of only two seconds; (c) a course correction was not necessary or even contemplated by the pilot; and (d) there was no corroborating evidence and the pilot himself even mentions considering it could have been a military jet. When used to update our belief regarding whether this was a drone incident that placed a commercial airliner at serious risk of harm, this fact scenario would update our belief in the following ways. *First*: that this was *probably not a drone*; and *second*: that *a collision was unlikely* to have resulted. Using this approach we would resolve that the airliner was not at any significantly increased risk of harm, yet this report (Airprox Report 2017035) was given a risk rating of B[87] and considered by the Airprox board to be one where safety was not assured - resulting in headlines proclaiming the incident a *serious near miss* and promoting irrational fears of drone batteries exploding in mid-air and bringing down commercial airliners (Tonkin, 2017).

We turn to another fact scenario where: (a) the airliner was at 4300 feet on approach[88] to Heathrow airport; (b) only one (co-)pilot observed the object claimed to be a drone that he described as 2 metres (almost 7 feet) across[89], and located forward and to the right of the cockpit in the one-o'clock position; (c) a course correction was not considered necessary or contemplated by the pilot; and (d) there had been two previous reports of 'a drone' in the area fifteen minutes prior and at different altitudes - making this the third report to air traffic control in a short window of time. Initially this report (Airprox Report 2016247) was rated as medium to high risk. However, a fourth pilot later advised that he had also observed the object and believed it not to be a drone - rather, he described it as a small bunch of white helium balloons. However, while the Airprox Board's final report downgraded the degree of risk to a C, they incredibly maintained in their report that *there was no doubt as to the object's identity as a drone* and described it as a *drone flown in conflict with a Boeing 777*. Bizarrely, one-third of all 2017 Airprox reports were given the highest risk rating of A - sometimes on significantly less risk-indicative detail. The reason we mostly focus on reports of events from 2017 in the footnotes of this work is because others have already expended considerable effort on similar analysis to identify deficiencies in Airprox reports from 2018-2020[90].

---

[85] There is one example of an actual drone collision with a small 6–7-seater commercial Beech King Air 100 while on approach to Jean Lesage Airport in Canada in October 2017. Contrary to the hyperbolic suggestions that a drone hitting an airplane was going to cause massive and potentially catastrophic damage, that plane, registration C-GJBV, landed safely and was back in service the following day. If a drone was going to bring down an airplane, this would have been the type of aircraft expected to have suffered extensively. (https://www.smithsonianmag.com/air-space-magazine/airliner-probably-hits-drone-180965320/).

[86] E.g.: Pilots of other aircraft or ground-based observers in the vicinity.

[87] Where the Airprox Board applied their own internally developed risk scale shown in Table 1.

[88] Being *on approach* means coming in for a landing - which is one of the highest workload periods for a pilot.

[89] Which is considerably large for what the pilot was describing as a quad-rotor drone. This author was unable to locate a white consumer quad-rotor drone larger than 3.4 feet across. Two manufacturers sold larger drones that reached 5 feet in width but were still smaller than the 7 feet described by the pilot. However, these were (a) quite expensive (over £10,000) and (b) were six- and eight-rotors respectively and not likely to be confused for a quad-rotor device.

Overscoring risk assessments as a result of using non-causal, limited and static risk profile scales like the one shown in Table 3 affects the way incidents go on to be described in regulator's reports[91], and leaves them open to further embellishment by the mainstream media through hyperbolic headlines portending airline catastrophe. While new regulation like EU 2019/947 and CAP722 promote the regulations' intentions toward risk identification and mitigation[92], when their risk position is based on both negatively biased risk assessments and public sentiment born out of fearmongering headlines the result becomes overly restrictive, overcomplicated and may be entirely unnecessary. Further, through integration of the proposed changes from Opinion No. 05/2019, the EU 2019/947 regulation introduces yet another non-causal, limited and static risk classification system that uses metrics about the drone as proxy for evaluating the *unmitigated risk of a person being hit* in a variety of operational scenarios[93]. While the drone's weight has a direct bearing on the outcome, the modern sensor and AI assisted drone model's width or weight are generally not causal in the accident itself. *Human factors* of the drone pilot (skill and acumen, situational awareness, decision-making, communication, threat and error management etc.) and the *environmental conditions* in which the drone is being flown (high winds, inclement weather etc.) normally have more causal influence on the potential for accident and should be far more prominently accounted for in any assessment of the potential risk drones pose to other aircraft, people and property.

A final issue is that the way in which these EU-devised drone regulations are being brought into effect serves to create revenue streams for drone manufacturers, training organisations, flying clubs, the CAA and councils[94] while intensifying the *hostile environment* for drone pilots.

---

[90] https://www.airproxrealitycheck.org

[91] The text shown in Table 3 for each risk level is regularly repeated, often verbatim, as the prevailing conclusion in UK Airprox Reports.

[92] EU 2019/947 is a one-size-fits-all blunt instrument that intermingles risks to safety (Section 3), privacy (Section 16), security (Sections 20-21) and the environment (Sections 20-21, 25) in a single strategy. It is largely built around the belief that adoption of the regulation will instantly ensure regulators and pilots of manned and unmanned aircraft will adhere to rules that set *risk level criteria* (Section 6) and identify *risks mitigation requirements* (Section 7) that are *proportionate to the nature and risk of the operation or activity* (Section 5). While EU 2019/947 is very prescriptive with regards to rules around documentation and reporting processes it is largely silent as to how to measure and mitigate risk.

[93] Opinion 01/2019, Appendix 1: Risk assessment for STS-01.

[94] Opinion No. 05/2019 provides that the proposed changes brought about by adoption of EU 2019/947 *will increase the cost-effectiveness for UAS operators, manufacturers and competent authorities.* However, this opinion's proposes changes (now part of EU 2019/947) which the author claims are intended to *address emerging safety issues,* further increasing restrictions and documentation requirements for drone pilots.

# 6. The future for drones and drone pilots: A synthesis of emerging technology, regulations and public policy

With CAP 722's transition period ending in December 2022, drone manufacturers will sell more new drones to drone enthusiasts who, after December 2022, are required to fly a *CE-Class marked* unit. The price for these units will likely increase because there is a process and fee for the manufacturer to attain CE-Class certification[95], along with the cost of any additional required technology not currently available in an existing drone model[96]. The majority of changes to current drone models consist of software updates and addition of the CE-Class mark on the base of the drone. In many cases it would be possible to retrospectively grant CE-Class mark certification to drones manufactured in the last 2-3 years, and while they have not presently done so some manufacturers have apparently suggested they could do this[97]. However, the CAA are yet to confirm whether the language of CAP722 is intended to require that drones are designed and visibly labelled with their CE-Class mark solely from manufacture[98]. While existing unmarked drones will continue to be allowed to fly, most will be placed under stricter provisions than those they currently fly under. In some cases these restrictions significantly reduce where and how they may be used - especially for the recreational user. As a result of weighing over 2kg, tens of thousands of recreational drone enthusiasts who own units like current model DJI Inspire and Phantom drones will find their options for flying from public open and green spaces become significantly limited[99]. Most people flying these legacy drones[100] will require additional certification[101] in order to continue to be legal when flying the same drones they have been flying safely, in most cases for several years.

The number of training establishments and their student capacities are already both increasing as they prepare to deal with the expected increase in drone enthusiasts who, under the new EU-based regulations, will be required to hold expensive A2 Certificate of Competency (CofC)[102] and General Visual Line of Sight Certificate (GVC)[103] endorsements in order to meet the updated drone pilot certification requirements.

The CAA and local authorities will see increased revenue from fees to assess and certify drones for CE-Class markings, and an increased number of drone enthusiasts applying for certification, permits and filming authorisations in order to simply fly their drone in the generally safe and responsible manner the majority have demonstrated for the last several years.

---

[95] The introduction of CE-class marking is described in CAP722: https://publicapps.caa.co.uk/cap722

[96] These features are mostly software-based. The radio transmitter will be required to broadcast the drone's remote identification. Drones will require ABS-B to sense other air traffic nearby, software for air avoidance, geofencing and the ability for smart return to home (to return to the location it took off from). Many existing drones already have these features.

[97] https://www.heliguy.com/blogs/posts/dji-drones-retrospective-ce-class-marking

[98] While CAP722 at section 2.2.1.3 states It is most important to note that an unmanned aircraft product can only be allocated within a UAS Class if it has been manufactured to the relevant product standard, independently assessed as being compliant, and visibly labelled as such, it has been suggested that the CAA may not have intended this to allow retrospective certification of drones manufactured before the requirement was drafted.

[99] The classification for drones over 2kg but less than 25kg will be equivalent to C3 and C4. C3 where the unit possesses automatic control modes, remote identification and geoawareness. C4 where it possesses no automated control systems. In both categories the drone can only be used *far from people*, meaning that the drone cannot be flown in any area where uninvolved people may be present in the flight zone.

[100] Legacy drones is the term used in CAP722 to describe existing drones that do not have CE-Class marking.

[101] These requirements are A2 CofC and GVC certifications.

[102] The average cost of A2 CofC training for certification is between £250-295.

[103] GVC licensing requires 1-3 days of classroom training and in-person flight tests, costing between £600 and £950 - and in some cases, more than the cost of the drone the recreational pilot will go on to fly.

However, for many recreational drone pilots it will seem easier either to give up drone flying altogether, or at the very least to dispose of existing serviceable drones and batteries and replace them with new ostensibly identical but CE-Class marked drones. Either way, our government's assent to yet another grandfathered-in EU regulation serves only to increase the number of toxic and non-biodegradable substances in UK landfills - both from the plastics and circuit boards of these unmarked but serviceable drone units, and the contents of their rechargeable battery packs[104]. The last couple of years have taken so much from so many people. Overregulation, inconsistent regulation and disharmony need not add to their suffering.

While adoption of the new EU drone regulation that was the impetus for the UK's CAP722 mandates the presence of certain autonomous abilities[105], at present it does not regulate the design process, integration, or security of those elements of the drone's system. Already, we see drones that use rules engines, ML algorithms and technology that, like that seen in modern airliners and semi-autonomous motor vehicles, goes beyond *inertial navigation system* (INS)[106] to identify the current location of the craft through incorporating real-time *global positioning system* (GPS) data - resulting in GPS-aided INS (GPS-INS) that, along with visual, sonar and radar sensors, affords the drone an ability to locate and identify objects and maintain awareness of potential hazards and other aircraft that it may come into conflict with (Daley, 2019; Kinaneva et al, 2019; SESAR, 2020)[107]. An issue that may arise in future is whether, as we now see with autonomous cars, drones may attract (or necessitate) additional regulation that oversees the development, security and testing of these integrated autonomous features into their hardware and software[108].

Recommendations for future work should include: (i) evaluation of the strengths and limitations of existing aviation risk estimation tools that are applied to drones, such as the ICAO Risk Rating tool, new EU 2019/947 Ground Risk Classification, and the self-developed approach presently in use by the UK Airprox Board to grade drone proximity incidents; (ii) development and evaluation of a new causal tool for assessment and grading of existing incidents to provide known priors; for (iii) an approach for more accurate estimation of the probability of serious event outcomes given the established uncertain accuracy of eye-witness descriptions; and (iv) a model for dynamic estimation of potential risk that can be applied in near real-time for future events.

While our government and local authorities might like them to, drones are not going away - and they should not simply be relegated to the landfill as a result of inappropriately motivated, poorly couched or hurriedly enacted rules, hyperbolic headlines, and misconceptions about the true risk profile of the activity being regulated. Adoption of EU drone regulation without, for example, initiating the process to have the UK's own qualification and certification scheme recognised on parity with EU member states means the UK effectively gets the worst of both worlds: more demanding EU rules, but without the benefit of pre-emption of additional regulation on the national and subnational level as would apply in EU member states. This results in confusion and, as we are seeing, layers of regulatory response that operate collectively as dysfunctional public policy - creating a hostile environment for people engaged in the *now controlled* activity. Ensuring citizens stay within the bounds of law intended to guarantee public safety shouldn't be simply yet another case of *here comes the fun police*[109]. The easiest way remains making law *comprehensible*, *consistent* and *easy to comply with*. The existing framework for drone use is anything but.

---

[104] Most owners will possess multiple Li-Ion or Li-Polymer battery packs for each of their existing drones.

[105] Such as technology that enables the drone to return to home, or for object detection and avoidance to prevent conflict with other aircraft.

[106] Or as it is known in the drone world *real time kinematic* (RTK) positioning - which uses gyroscopes, accelerometers and in some cases, magnetometers to track the drone's position in flight relative to the base station or *Home* point.

[107] Commercially available units include the DJI Manifold and Matrox 300 RTK, both of which are capable of advanced computing, automated identification and analysis of targets of interest, smart tracking and search and rescue.

[108] For example, UN-ECE World Forum for Harmonisation of Vehicle Regulations (WP.29) requires vehicle manufacturers to consider the issues of cybersecurity throughout their entire design, manufacture, sales and post-marketing processes and throughout their component supply chain. The EU recently released the latest draft of (EU) 2019/2144 - their type-approval regulation for vehicles with Autonomous Driving Systems (ADS).

**Contribution Statement**
SM performed the primary research with the support of two research assistants. SM wrote the first draft. KD reviewed the technology sections. BS reviewed and informed the legal analysis. NF refined and extended the analysis and discussion of risk assessment. All authors reviewed and accepted the final manuscript.

**Acknowledgements**
The authors acknowledge support for this research from the following sources: RAEng under project *SafeAIR: Safer aviation from ethical Autonomous Intelligence Regulation* - ICRF2122-5-234; EPSRC under project *AISEC: AI Secure and explainable by Construction* - EP/T026952/1; and UKRP grant EP/V026607/1 - Trustworthy Autonomous Systems Node in Governance and Regulation. The authors also acknowledge the School of Law at LJMU for funding our two research assistants' efforts on this work.

The authors acknowledge the assistance of research assistants George Lamb and Majida Ismael from the School of Law at Liverpool John Moores University who each provided around 80 hours of data collection and preliminary analysis.

**Conflict Statement**
No author reported a conflict of interest relevant to this work.

---

[109] The Macmillan Dictionary defines *fun police* as: *people who want to stop other people having fun*. Last accessed: 10 March 2022. Sourced from: https://www.macmillandictionary.com/dictionary/british/fun-police

# Appendix A: Local Authorities reviewed in this work

| District Councils | In Figures 3 & 4 | Blanket Ban |
|---|---|---|
| Adur District Council | | |
| Allerdale Borough Council | | |
| Amber Valley Borough Council | | |
| Arun District Council | | |
| Ashfield District Council | | |
| Ashford Borough Council | | |
| Aylesbury Vale District Council | | X |
| Babergh District Council | | |
| Barrow-in-Furness Borough Council | | |
| Basildon Borough Council | | |
| Basingstoke & Deane Borough Council | | |
| Bassetlaw District Council | | |
| Blaby District Council | | |
| Bolsover District Council | | |
| Boston Borough Council | | |
| Braintree District Council | | |
| Breckland District Council | | |
| Brentwood Borough Council | | |
| Broadland District Council | | |
| Bromsgrove District Council | | |
| Broxbourne Borough Council | | |
| Broxtowe Borough Council | | |
| Burnley Borough Council | | X |
| Cambridge City Council | | |
| Cannock Chase District Council | | |
| Canterbury City Council | | |
| Carlisle City Council | | |
| Castle Point District Council | | |
| Charnwood Borough Council | | |
| Chelmsford City Council | | X |
| Cheltenham Borough Council | | |
| Cherwell District Council | | |
| Chesterfield Borough Council | | |
| Chichester District Council | | |
| Chiltern District Council | | X |
| Chorley Borough Council | | |
| Colchester Borough Council | | X |
| Copeland Borough Council | | |
| Corby Borough Council | | |
| Cotswold District Council | | |
| Craven District Council | | |
| Crawley Borough Council | | |
| Dacorum Borough Council | | |
| Dartford Borough Council | | X |
| Daventry District Council | | X |
| Derbyshire Dales District Council | | X |
| Dover District Council | | X |
| East Cambridgeshire District Council | | X |
| East Devon District Council | | |
| East Hampshire District Council | | |
| East Hertfordshire District Council | X | X |
| East Lindsey District Council | | |
| East Northamptonshire District Council | | |
| East Staffordshire Borough Council | | X |
| East Suffolk Council | | |
| Eastborne Borough Council | | |
| Eastleigh Borough Council | | |
| Eden District Council | | |
| Elmbridge Borough Council | | |
| Epping Forest District Council | | |
| Epsom & Ewell Borough Council | | |
| Erewash Borough Council | | |
| Exeter City Council | | |
| Fareham Borough Council | | |
| Fenland District Council | | |
| Folkestone & Hythe District Council | | |
| Forest of Dean District Council | | |
| Fylde Borough Council | | |
| Gelding Borough Council | | |
| Gloucester City Council | | |
| Gosport Borough Council | | |
| Gravesham Borough Council | | |
| Great Yarmouth Borough Council | | |
| Guildford Borough Council | | |
| Hambleton District Council | | |
| Harborough District Council | | |
| Harlow District Council | | |
| Harrogate Borough Council | | |
| Hart District Council | | |
| Hastings Borough Council | | |
| Havant Borough Council | | |
| Hertsmere Borough Council | | |
| High Peak Borough Council | | |
| Hinkley & Bosworth Borough Council | | X |
| Horsham District Council | | |
| Huntingdonshire District Council | | |
| Hyndburn Borough Council | | |
| Ipswich Borough Council | | |
| Kettering Borough Council | | |
| Kings Lynn & West Norfolk Borough Council | | |
| Lancaster City Council | | |
| Lewes District Council | | |
| Lichfield City Council | | X |
| Lincoln City Council | | |
| Maidstone Borough Council | | |
| Maldon District Council | | |
| Malvern Hills District Council | | X |
| Mansfield District Council | | |
| Melton Borough Council | | |
| Mendip District Council | | |
| Mid Devon District Council | | |
| Mid Suffolk District Council | | |
| Mid Sussex District Council | | |
| Mole Valley District Council | | |
| North Devon District Council | | |
| North East Derbyshire District Council | | |
| North Hertfordshire District Council | | |
| North Kesteven District Council | | |
| North Norfolk District Council | | |
| North West Leicestershire District Council | | |
| North Warwickshire District Council | | |
| New Forest District Council | | X |
| Newark & Sherwood District Council | | |
| Newcastle-Under-Lyme Borough Council | | |
| West Northamptonshire Council | | |
| Norwich City Council | | |
| Nuneaton & Bedworth Borough Council | | |
| Oadby & Wigston Borough Council | | X |
| Oxford City Council | | |
| Pendle Borough Council | | X |
| Preston City Council | | |
| Redditch Borough Council | | X |
| Reigate & Banstead Borough Council | | X |
| Ribble Valley Borough Council | | |
| Richmondshire District Council | | |
| Rochford District Council | | |
| Rossendale Borough Council | | X |
| Rother District Council | | |
| Rugby Borough Council | | X |
| Runnymede Borough Council | | |
| Rushcliffe Borough Council | | |
| Rushmoor Borough Council | | X |
| Ryedale District Council | | |
| Somerset West & Taunton Council | | |
| South Bucks District Council | | X |
| South Cambridgeshire District Council | | |
| South Derbyshire District Council | | |
| South Hams District Council | | |
| South Holland District Council | | |
| South Kesteven District Council | | |
| South Lakeland District Council | | |
| South Norfolk District Council | | X |
| South Hamptonshire District Council | | |
| South Oxfordshire District Council | | |
| South Ribble Borough Council | | X |
| South Somerset District Council | | |
| South Staffordshire District Council | | |
| Scarborough Borough Council | | |
| Sedgemoor District Council | | |
| Selby District Council | | |
| Sevenoaks District Council | | |
| Spelthorne Borough Council | | |
| St Albans City Council | | |
| Stafford Borough Council | | |
| Staffordshire Moorlands District Council | | |
| Stevenage Borough Council | | X |
| Stafford on Avon District Council | | |
| Stroud District Council | | |
| Suffolk Coastal District Council | | |
| Surrey Heath Borough Council | | X |
| Swale Borough Council | | X |
| Tamworth Borough Council | | |
| Tandridge District Council | | |
| Teignbridge District Council | | |
| Tendring District Council | | |
| Test Valley Borough Council | | |
| Tewkesbury Borough Council | | |
| Thanet District Council | | |
| Three Rivers District Council | | |
| Tonbridge & Malling Borough Council | | X |
| Torridge District Council | | |
| Tunbridge Wells Borough Council | | |
| Uttlesford District Council | | |
| Vale of White Horse District Council | | |
| Warwick District Council | | |
| Watford Borough Council | | X |
| Waverley Borough Council | | |
| Wealden District Council | | |
| Wellingborough Borough Council | | |
| Welwyn Hatfield Borough Council | | |

| | | In Figures 3 & 4 | Blanket Ban |
|---|---|---|---|
| | West Devon District Council | | X |
| | West Lancashire District Council | | |
| | West Lindsey District Council | | |
| | West Oxfordshire District Council | | |
| | West Suffolk Council | | X |
| | Winchester City Council | | |
| | Woking Borough Council | | |
| | Worcester City Council | | |
| | Worthing Borough Council | | |
| | Wychavon District Council | | X |
| | Wycombe District Council | | X |
| | Wyre Borough Council | | |
| | Wyre Forest District Council | | X |

| | | In Figures 3 & 4 | Blanket Ban |
|---|---|---|---|
| **County Councils** | | | |
| | Buckinghamshire County Council | X | X |
| | Cambridgeshire County Council | X | |
| | Cumbria County Council | X | |
| | Derbyshire County Council | X | X |
| | Devon County Council | X | X |
| | East Sussex County Council | X | |
| | Essex County Council | X | |
| | Gloucestershire County Council | X | X |
| | Hampshire County Council | X | |
| | Hertfordshire County Council | X | |
| | Kent County Council | X | |
| | Lancashire County Council | X | |
| | Leicestershire County Council | X | |
| | Lincolnshire County Council | X | |
| | Norfolk County Council | X | |
| | North Yorkshire County Council | X | |
| | Northamptonshire County Council | X | |
| | Nottinghamshire County Council | X | |
| | Oxfordshire County Council | X | |
| | Somerset County Council | X | |
| | Staffordshire County Council | X | X |
| | Suffolk County Council | X | |
| | Surrey County Council | X | |
| | Warwickshire County Council | X | X |
| | West Sussex County Council | X | X |
| | Worcestershire County Council | X | |

| | | In Figures 3 & 4 | Blanket Ban |
|---|---|---|---|
| **Unitary Authorities** | | | |
| | Bath & North East Somerset Council | X | |
| | Bedford Borough Council | X | |
| | Blackburn with Darwen Borough Council | X | |
| | Blackpool Council | | |

| | | |
|---|:---:|:---:|
| Bournemouth, Christchurch & Poole Council | X | |
| Bracknell Forest Borough Council | | |
| Brighton & Hove City Council | X | |
| Bristol City Council | X | |
| Central Bedfordshire Council | X | |
| Cheshire East Council | X | |
| Cheshire West and Chester Council | X | |
| Cornwall Council | X | |
| Durham County Council | X | |
| Darlington Borough Council | X | |
| Derby City Council | X | |
| Dorset Council | X | |
| East Riding of Yorkshire Council | X | |
| Halton Borough Council | X | |
| Hartlepool Borough Council | X | |
| Herefordshire Council | X | |
| Isle of Wight Council | X | |
| Hull City Council | | |
| Leicester City Council | X | |
| Luton Borough Council | X | X |
| Medway Council | X | |
| Middlesbrough Borough Council | X | |
| Milton Keynes Council | X | |
| North East Lincolnshire Council | X | |
| North Lincolnshire Council | X | |
| North Somerset Council | X | |
| Northumberland County Council | X | |
| Nottingham City Council | X | |
| Peterborough City Council | X | |
| Plymouth City Council | X | |
| Portsmouth City Council | X | |
| Reading Borough Council | | |
| Redcar & Cleveland Borough Council | | |
| Rutland County Council | X | |
| Shropshire Council | X | |
| Slough Borough Council | | |
| Southampton City Council | X | |
| Southend-on-Sea Borough Council | | |
| South Gloucestershire Council | X | |
| Stockton-on-Tees Borough Council | X | |
| Stoke-on-Trent City Council | X | |
| Swindon Borough Council | X | |
| Telford & Wrekin Borough Council | X | |
| Thurrock Burrough Council | X | |
| Torbay Council | | |
| Warrington Borough Council | X | |
| West Berkshire Council | X | |
| Wiltshire Council | X | |
| Windsor & Maidenhead Borough Council | | |
| Wokingham Borough Council | | |
| City of York Council | X | |

| | In Figures 3 & 4 | Blanket Ban |
|---|:---:|:---:|
| **Metropolitan Districts** | | |
| Barnsley Borough Council | X | |
| Birmingham City Council | X | X |
| Bolton Borough Council | | |
| Bradford City Council | X | |

| | | | |
|---|---|---|---|
| | Bury Borough Council | | |
| | Calderdale Borough Council | X | |
| | Coventry City Council | X | |
| | Doncaster Borough Council | X | |
| | Dudley Borough Council | X | |
| | Gateshead Borough Council | X | |
| | Kirklees Borough Council | X | |
| | Knowsley Borough Council | X | |
| | Leeds City Council | | X |
| | Liverpool City Council | X | |
| | Manchester City Council | X | X |
| | North Tyneside Borough Council | X | |
| | Newcastle Upon Tyne City Council | X | X |
| | Oldham Borough Council | | |
| | Rochdale Borough Council | | |
| | Rotherham Borough Council | X | |
| | South Tyneside Borough Council | X | |
| | Salford City Council | | |
| | Sandwell Metropolitan Council | X | |
| | Sefton Borough Council | X | |
| | Sheffield City Council | | |
| | Solihull Borough Council | X | |
| | St Helens Borough Council | | |
| | Stockport Borough Council | | |
| | Sunderland City Council | X | |
| | Tameside Borough Council | | |
| | Trafford Borough Council | | |
| | Wakefield City Council | X | |
| | Walsall Borough Council | X | |
| | Wigan Borough Council | | |
| | Wirral Borough Council | X | |
| | Wolverhampton City Council | X | |

# Appendix B: Authority to make byelaws

A local authority such as a Parish Council has power to make byelaws in relation to the following matters:

| Function | Power |
|---|---|
| Regulating public walks or pleasure grounds provided by the council or to the cost of which the council has contributed | *Public Health Act* 1875, Section 164 |
| Regulating the letting for hire of pleasure boats in a park or pleasure ground provided or managed by the council | *Public Health Act* 1961, Section 54 |
| Regulating an open space or burial ground owned or controlled by the council | *Open Spaces Act* 1906, Section 15 |
| Managing mortuaries and post-mortem rooms provided by the council | *Public Health Act* 1936, Section 198 |
| Regulating baths, washhouses, swimming baths and bathing places under the council's management | *Public Health Act* 1936, Section 223 |
| Regulating public bathing in the area | *Public Health Act* 1936, Section 231 |
| Regulating swimming baths and bathing places not managed by the council and which are open to the public at a charge | *Public Health Act* 1936, Section 233 |
| Regulating parking places for bicycles and motor cycles provided by the council | *Road Traffic Regulation Act* 1984, Section 57(7) |
| Regulating markets | *Food Act* 1984, Section 60 |